\documentclass[sn-mathphys,Numbered]{sn-jnl}% Math and Physical Sciences Reference Style
\usepackage{graphicx}%
\usepackage{multirow}%
\usepackage{amsmath,amssymb,amsfonts}%
\usepackage{amsthm}%
\usepackage{mathrsfs}%
\usepackage[title]{appendix}%
\usepackage{xcolor}%
\usepackage{textcomp}%
\usepackage{manyfoot}%
\usepackage{booktabs}%
\usepackage{algorithm}%
\usepackage{algorithmicx}%
\usepackage{algpseudocode}%
\usepackage{listings}%
\theoremstyle{thmstyleone}%
\theoremstyle{thmstyletwo}%

\theoremstyle{thmstylethree}%

\begin{document}

\title[Article Title]{Thermodynamic formalism and anomalous transport in 1D semiclassical Bose-Hubbard chain}


\author[1,2]{\fnm{Dragan} \sur{Marković}}\email{vokramnagard@gmail.com} 

\author[2]{\fnm{Mihailo} \sur{Čubrović}}\email{cubrovic@ipb.ac.rs}

\affil[1]{\orgdiv{Department of Physics}, \orgname{University of Belgrade}, \orgaddress{\street{Studentski Trg 12-16}, \city{Belgrade}, \postcode{11000}, \country{Serbia}}}

\affil[2]{\orgdiv{Center for the Study of Complex Systems}, \orgname{Institute of Physics Belgrade}, \orgaddress{\street{Pregrevica 118}, \city{Belgrade}, \postcode{11080}, \country{Serbia}}}


\abstract{We analyze the time-dependent free energy functionals of the semiclassical one-dimensional Bose-Hubbard chain. We first review the weakly chaotic dynamics and the consequent early-time anomalous diffusion in the system. The anomalous diffusion is robust, appears with strictly quantized coefficients, and persists even for very long chains (more than hundred sites), crossing over to normal diffusion at late times. We identify fast (phase) and slow (number) variables and thus consider annealed and quenched partition functions, corresponding to fixing the number variables or integrating over them, respectively. We observe the leading quantum effects in the annealed free energy, whereas the quenched energy is undefined in the thermodynamic limit, signaling the absence of thermodynamic equilibrium in the quenched regime. But already the leading correction away from the quenched regime reproduces the annealed partition function exactly. This encapsulates the fact that in both slow- and fast-chaos regime both the anomalous and the normal diffusion can be seen (though at different times).}




\maketitle

\newpage

\section{Introduction}\label{sec1}

The vast and interesting phenomenology of cold atom systems \cite{ColdRev}, the universal behavior of theoretical models such as the SYK model \cite{SachdevYe,MaldacenaSYK,VandorenSYK}, and novel indicators of quantum dynamics such as OTOC \cite{BlackButter,MSSbound,ScaffidiScrChaos,HashimotoqmOTOC} and Krylov complexity \cite{Roberts:2016hpo,Jefferson:2017sdb,Rabinovici:2020,Caputa:2021} have led to resurgent interest in quantum chaos. The Bose-Hubbard model is an example of a cold-atom system which is nonintegrable and exhibits quantum chaos \cite{Kolovsky:2004,Kollath:2010,Kolovsky:2016,Pausch2body,Richaud_2018,Ferrari:2023zfe,markovic2023chaos}. A convenient property of this model is also that it has a classical limit \cite{Polkovnik2003Main,Nakerst:2022prc}, facilitating the comparisons of classical and quantum dynamics. A natural question in this context is how (and if at all) the chaotic dynamics leads to thermalization and hydrodynamics (normal transport). Few detailed studies on this matter exist for the Bose-Hubbard model, which motivates our study.

More specifically, our goal is to understand the interplay between chaos, the transport which we have previously found to be strongly anomalous \cite{markovic2023chaos}, and the thermodynamic functions of the system. Anomalous transport is expected in weakly chaotic systems \cite{ZASLAVSKY2002461,zaslavsky2007physics}, however in our case it has strictly integer exponents $2m$ and $4m$, where $m$ is a non-negative integer. This is surprising as the anomalous exponents are usually fractional \cite{zaslavsky2007physics}. We have found that anomalous diffusion holds even for enormous chains, with $L>100$ sites. This is also surprising as we expect the relative measure of stable regions in phase space to diminish to zero in the $L\to\infty$ limit, leading to strong chaos and normal diffusion. This set of issues is also of relevance for the studies of chaos control in quantum many-body systems \cite{Andreev:2019arXiv190703602A}, and for a broader understanding of hyperchaotic systems \cite{McCormack:2021Photo...8..554M}, as our system also displays hyperchaos \cite{markovic2023chaos}. The former issue (chaos control) is also important for possible realization of Bose-Hubbard chains in quantum computing \cite{npjQI...6...58Y,Zhang:2023vtu}.

In this work we look at weak chaos and anomalous transport from the thermodynamic point of view: we calculate the partition function and free energy of the system. Since the dynamics exhibits timescale separation into fast and slow variables, it is natural to consider two possible definitions of the partition function: integrate over both fast and slow variables (annealed partition function) or solely over the fast variables, fixing the slow ones (quenched partition function). With some hindsight, we can say that the two approaches correspond to different epochs in the evolution of the system: early-time anomalous diffusion versus long-time normal diffusion regime. Nevertheless, the complete picture is still evasive, as we can only evaluate the free energies at leading order, in a very crude approximation.

\section{The model}\label{sec2}

We consider a one-dimensional Bose-Hubbard chain with $L$ sites, whose Hamiltonian is given by:
\begin{equation}
\label{bh}H_\mathrm{BH}=\sum_{j=1}^L\left[-J\left(b_j^\dagger b_{j+1}+b_jb^\dagger_{j+1}\right)+\frac{U_\mathrm{BH}}{2}n_j\left(n_j-1\right)-\mu n_j\right],
\end{equation}
where $J$ is the hopping parameter, $U_{BH}$ is the on-site Coulomb repulsion, $b_j^\dagger,b_j$ are bosonic creation and annihilation operators, $n_j$ is the occupation number of the site $j$ and $\mu$ is the chemical potential. We do not impose periodic boundary conditions, so $b_j^\dagger,b_j \equiv 0$ for $j=0,L+1$. 

We are interested in the semiclassical regime, i.e. when the number of particles $N=\sum_j n_j$ goes to infinity while the number of sites $L$ stays fixed. The semiclassical Hamiltonian is obtained by introducing new variables: $(b_j^\dagger,b_j)\mapsto (\psi_j^*,\psi_j)\equiv(b_j^\dagger,b_j)/\sqrt{N}$ \cite{Polkovnik2003Main,Nakerst:2022prc,markovic2023chaos}. The commutator of new variables vanishes as $1/N$ and the Hamiltonian becomes classical, with rescaled Coulomb repulsion parameter $U\equiv NU_\mathrm{BH}$ and the number-conservation constraint $\sum_j |\psi_j|^2=1$.
%\begin{equation}
%\label{bhclass}H\equiv\lim_{N\to\infty}\frac{1}{N}H_\mathrm{BH}=\sum_{j=1}^L\left[-\frac{J}{2}\left(\psi_j^*\psi_{j+1}+\psi_j\psi^*_{j+1}\right)+\frac{U}{2}\vert\psi_j\vert^4-\mu\vert\psi_j\vert^2\right].
%\end{equation}
% At this point it is useful to introduce two new sets of variables: $P,Q$ coordinates given as re scaled real and imaginary parts of the $\psi$ variables:
%\begin{equation}
%\label{pqtrans}Q_j=\frac{i}{\sqrt 2}(\psi_j-\psi^*_j),~P_j=\frac{1}{\sqrt 2}(\psi_j+\psi^*_j).
%\end{equation}
%These variables are useful for numerical calculations, which we utilize heavily in our work and for the introduction of action angle variables, which are used for both numerical and analytical treatment. In relationship with the $P,Q$ variables they are given as:
%\begin{equation}
%\label{actangle}P_j=\sqrt{2I_j}\sin\phi_j,~~Q_j=\sqrt{2I_j}\cos\phi_j.
%\end{equation}
Finally, we introduce the number-phase variables $(I_j,\phi_j)$. In the integrable limit $J\to 0$ the Hamiltonian will only depend on $I_j$ so the number-phase variables become action-angle variables: the actions are integrals of motion and the angles have periodic dynamics. In the general, chaotic case, strictly speaking there are no action-angle variables, although in the literature one still often keeps the same terminology so that the "actions" are really the quasiintegrals of motion which evolve slowly whilst the "angles" change rapidly and play the role of "fast" variables that we can average over. We will nevertheless mostly use the number-phase terminology in order to remain precise. In these variables the Hamiltonian reads:
\begin{equation}
\label{bhclassiphi}H=\sum_{j=1}^L{}\left(\frac{U}{2}I_j^2-\mu I_j\right)-2J\sum_{j=1}^{L-1}{}\sqrt{I_jI_{j+1}}\cos\left(\phi_j-\phi_{j+1}\right),
\end{equation}
Since $I_j$ are the occupation numbers for each site, they satisfy the constraint $\sum_j I_j=1$. From Eq.~(\ref{bhclassiphi}) we arrive at the equations of motion:
\begin{eqnarray}
&&\dot{\phi_j}=-\mu+UI_j-J\left(\sqrt{\frac{I_{j-1}}{I_j}}\cos\left(\phi_j-\phi_{j-1}\right)+\sqrt{\frac{I_{j+1}}{I_j}}\cos\left(\phi_j-\phi_{j+1}\right)\right)\label{eomi}\\
&&\dot{I_j}=2J\left(\sqrt{I_jI_{j-1}}\sin\left(\phi_{j-1}-\phi_j\right)+\sqrt{I_jI_{j+1}}\sin\left(\phi_{j+1}-\phi_j\right)\right)\label{eomphi}.
\end{eqnarray}
Equations of motions are nonlinear as they have to be in a nonintegrable system.\footnote{Of course, (non)linearity of the equations of motion is in general a matter of specific generalized coordinates and momenta used. Integrable systems can also have nonlinear equations of motion for some choices of canonical variables. But the Bose-Hubbard model is known to be nonintegrable as it exhibits chaos, which we also find when computing the Lyapunov exponents. Consequently, its (exact) equations of motion cannot be linearized by any well-behaving canonical transformation.} In special cases $J=0$ and $U=0$ the system becomes integrable. In general, low/high $U/J$ ratio corresponds to tight/weak binding regime, leading to the superfluid and Mott insulator regimes respectively \cite{PolkovnikSachdev2002,Polkovnik2003Main,Polkovnik2003}.\footnote{Since the model is one-dimensional there are no phase transitions but we can speak of two regimes separated by a crossover.}

\section{Chaos and transport}\label{sec3}

Exploring the semiclassical dynamics of the system it is appropriate to first calculate the Lyapunov exponents in order to characterize chaos. In \cite{markovic2023chaos} we have observed that the sites with a large initial occupation number in general have largest Lyapunov exponents which do not change significantly with the growth of $U$; initially empty sites have the lowest Lyapunov exponents (Figs. 4 and 5 in \cite{markovic2023chaos}). To further corroborate this we have also computed the Lyapunov exponent for varying $\mu/J$ and $U/J$ (Figs. 3, 4 and 5 in \cite{markovic2023chaos}), and found that, for strongly chaotic sites, the strength of chaos is almost independent of the system parameters. Initially filled sites show uniformly strong chaos, which suggests that chaos is driven by initial conditions (Fig. 3 in \cite{markovic2023chaos}).

We now move to the central result of our work so far. We consider a population of orbits in phase space with initial conditions distributed, e.g. as a Gaussian peaked at the point $(I_n(0),\phi_n(0))$. For strongly chaotic systems we expect to find diffusion in the space of numbers ("actions") \cite{Lichtenberg:1989,ZASLAVSKY2002461}. We inspect it by calculating:
\begin{equation}
\sigma^2(I_n(t))=\langle I_n^2(t)\rangle-\langle I_n(t)\rangle^2
\end{equation}
for the observed population of orbits. We observe strong superdiffusion, that is $\sigma^2\sim t^\zeta$ where $\zeta>1$. The only instances where this strongly anomalous diffusion is absent are the initially filled sites, which show no diffusion at all, i.e. they have $\zeta=0$. The coefficients are completely independent of the parameters of the system ($\mu/J$, $U/J$). They are thus a characteristic of the model and depend solely on the geometry of the initial conditions.

\begin{figure}[H] 
\centering
\includegraphics[width=.99\linewidth]{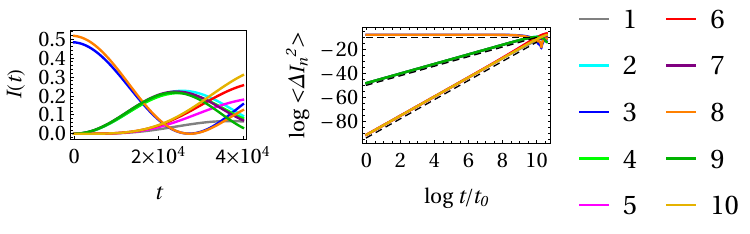}
\caption{Individual orbits $I_i(t)$ (left) and the log-log plot of the second central moment (variance) of the occupation numbers (right) for an ensemble of orbits in the chain of length $L=10$, with $U/J=0.375,\mu/J=0.25$. Initially the filled sites are $n=3,8$. The exponents take values $0$, $4$ or $8$, determined by the distance from the nearest initially occupied site. Black dashed lines are analytic plots $\langle\Delta I_n^2\rangle\propto t^{4m}$. The legend (far right) relates the site numbers to line colors.}
\label{figanomdiff1}
\end{figure}

In general the anomalous exponent is $4m$, with $m$ a positive integer which equals the distance of the given site from the nearest initially filled site. This also means that initially filled sites have the exponent $\zeta=0$ (as for a filled site the distance from the nearest filled site is zero). This is observed for almost all initial conditions, and for all ranges of parameters. In special cases, when the initial conditions are sufficiently complicated and many sites are initially partially filled with similar fillings, we also observe exponents equal to $2m$ with the same meaning for $m$ as before. In Fig.~\ref{figanomdiff1} we show the typical case, when the exponents equal $4m$.

%\begin{enumerate}
%\item{The anomalous transport exponents for the action $I_n$ take the values $\zeta_n=2m$ or $\zeta_n=4m$ (for $m=0,1,2,\ldots$) -- they are strictly even integers and come in arithmetic series with spacing $2$ or $4$. We find this very striking -- such integer and strictly "quantized" values are very rare in the literature \cite{Lichtenberg:1989}.}
%\item{The value of the exponent $\zeta_n$ for each site depends on the initial conditions, that is on the distribution of filled and empty sites. If the site $n$ is $m$ sites away from the nearest filled site $n_0$ (so that $m=\vert n-n_0\vert$), then the exponent is generically $\zeta_n=4m$.}
%\item{In particular, the filled site has $m=0$ -- the distance from the nearest filled site is zero, thus $\zeta_n=0$ -- the distribution function spreads at most logarithmically.}
%\item{If partially filled sites with similar occupation numbers are present (so that the notion of distance to the nearest filled site is vague), then the exponents take the form $2m$.}
%\item{The endpoints of the chain act effectively as filled sites, meaning that the relation $m=\vert n-n_0\vert$ is modified as $m=\mathrm{min}\left(\vert n-n_0\vert,n,L-n\right)$.}
%\end{enumerate}

%This is why the exemption periodic boundary conditions is so important: according to our observations one can simply arrive at the conclusion that on a Bose - Hubbard ring the notion of distance between sites is ambiguous, i.e. the distances are not unique. That is why the system equilibrates very quickly and does not show anomalous diffusion.

The origin of anomalous coefficients is very hard to understand \cite{METZLER20001,zaslavsky2007physics}, but according to Zaslavsky \cite{Zaslavsky1994FractionalKE} we can give a crude explanation, at least for the case $\zeta=4m$. For this case, we can use the non-resonant perturbation theory where the perturbed Hamiltonian has the form of a pendulum Hamiltonian \cite{markovic2023chaos}. It turns out that the period of oscillations of the pendulum is proportional to the square root of the number variable: $T\sim \sqrt{I_j}$. Rescaling these quantities by some factors $\lambda_T$ and $\lambda_I$, we find that the system stays invariant if $\lambda_T^2=\lambda_I$. Extending this to $m$ sites one derives: $\lambda_T^{2m}=\lambda_I$. From the Renormalization Group of kinetics formalism \cite{Zaslavsky1994FractionalKE,ZASLAVSKY2002461,zaslavsky2007physics}, the diffusion coefficients are given as:
\begin{equation}
    \zeta_m=\frac{2\log \lambda_I}{\log \lambda_T}=4m,
\end{equation}
just like in the numerics.

\begin{figure}[H] 
\centering
\includegraphics[width=.99\linewidth]{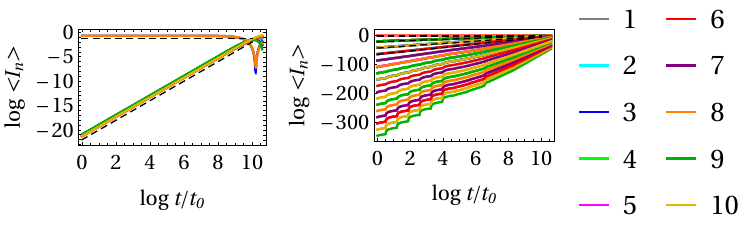}
\caption{The log-log plot of the first moment (mean) of the occupation numbers for an ensemble of orbits for two configurations with $U/J=0.375,\mu/J=0.25$. In the left panel we study the chain of length $L=10$ with sites $n=3,8$ initially filled; in the right panel we study the chain with $L=50$ where the sites $n=20,25,29$ are initially filled. The exponents take values $2m-2$ where $m$ is the distance from the nearest initially filled site (if $2m-2$ is not positive, then the exponent is simply zero). Black dashed lines are analytic plots $\langle I_n\rangle\propto t^{2m-2}$; for the $L=50$ chain we only plot a few lines with lowest exponents both in order not to cram the figure too much and also because the curves corresponding to very low occupation numbers can hardly be trusted because of numerical errors. The legend (far right) relates the site numbers to line colors in the left panel; in the right panel we use the same colors for site numbers modulo 10, i.e. each color stands for five different sites.}
\label{figanomdiff2}
\end{figure}

While anomalous transport is typically demonstrated by the spread of the distribution, i.e. the second moment, as we did in Fig.~\ref{figanomdiff1}, we can also study the first moment and its exponent $\xi$, given by $\langle I\rangle\sim t^\xi$. In Fig.~\ref{figanomdiff2} we plot the mean of the distribution on the logarithmic scale, and again find anomalous behavior with the exponent consistent with the hypothesis $\xi=\mathrm{max}(0,2m-2)$, where again $m$ is the distance to the nearest filled site. Notice that this in general means that the second moment does not scale as the square of the first moment. This is not surprising as the ensemble averaging and the square root usually do not commute. It confirms the robustness of anomalous transport in the Bose-Hubbard model.

%\begin{figure}[H] 
%\centering
%\includegraphics[width=.9\linewidth]{CrossToNormdiff.pdf}
%\caption{Evolution of the variance of the actions for an ensemble of orbits $\langle I_n^2(t)\rangle$ for the chain of length $L=10$, with $U/J=10$, $\mu/J=0$. At earlier times ($t/t_0\lesssim\exp(7)$) there is the usual robust anomalous diffusion, with exponents $\zeta=0,4,8$ (dashed lines in the left panel). After a fluctuating transitional period $\exp(7)\lesssim t/t_0\lesssim\exp(10)$, the initially non-filled sites exhibit normal diffusion with exponent $\zeta_n=1$, i.e. $\langle I_n^2(t)\propto t\rangle$ (dotted lines in the left panel and the zoom-in of the late-time epoch for $n=1,8$ in the right panel).}
%\label{fignormdiff}
%\end{figure}

Finally, everything we have said so far of anomalous transport remains true up to some (large) time $t_0$. Indeed we would expect that at some point the system approaches some form of thermal equilibrium, even if the time to reach it can be very large. This is indeed observed in our numerical calculations. In Fig. 11 (in \cite{markovic2023chaos}) we can follow how anomalous diffusion becomes normal, with $\zeta=1$, for sufficiently large times.

\subsection{Two-dimensional generalization}

\begin{figure}[H] 
\centering
\includegraphics[width=.99\linewidth]{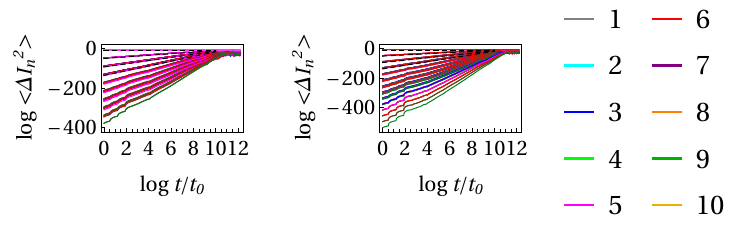}
\caption{Log-log plot of the second central moment (variance) of the number variables for an ensemble of orbits on a $10\times 10$ square lattice, with $U/J=50,\mu/J=0$. Initially filled sites are $(2,3)$ and $(7,4)$ (left) and $(2,3)$ and $(3,3)$ (right), and the color legend enumerates the sites 1 thru 10 in each row, i.e. the colors repeat 10 times in the whole $10\times 10$ model. The black dashed lines delineate the power laws $t^{4m}$, with $m=0,1,2,4$.}
\label{figanomdiff2d}
\end{figure}

We will now show that the same general conclusions, in particular the $4m$ series of exponents, remain valid also for a two-dimensional Bose-Hubbard model. We have no ambitions to go for a detailed analysis of the two-dimensional case which is known to behave very differently from the one-dimensional chain, e.g. in the sense that it has a sharp phase transition. For now we just want to demonstrate that the general behavior is not specifically tied to one-dimensional physics. Fig.~\ref{figanomdiff2d} makes the point: the same general law applies, 

\section{Thermodynamics}\label{sec4}

We have mentioned that at long times the system reaches the normal diffusion regime. We are still not sure about the meaning of the anomalous regime: is it a pre-thermalized regime or a regime which captures long-distance correlations and thus is not hydrodynamic. In order to shed some light on this, we calculate the partition functions of the system. The catch is that in principle we need to compute them as a function of time, i.e. for the evolving values of the variables -- this is necessary because our system is strongly out-of-equilibrium. Instead of integrating over all possible exact trajectories which is a hopeless task we will use two drastic approximations. Of course, it would be preferable to apply the usual formalism of non-equilibrium thermodynamics but for now we just want to have some indication of what is happening in the simplest possible approach.

The are now two possible limits: annealed and quenched. Annealed partition function is obtained when treating all the variables equally, that is, we integrate over the whole phase space, both numbers and phases. But since the phases are fast-winding variables and the numbers change slowly, it also make sense to consider the \emph{quenched} approximation where we freeze the numbers at their initial values and then integrate the logarithm of the partition function for the phases to arrive at the quenched free energy. 

\subsection{Annealed partition function}\label{subsec2}

By definition, the partition function is given as:

\begin{equation}
    Z_a=\int d\vec{I}d\vec{\phi}\exp\bigg(2\beta J\sum_{j=1}^{L-1}\sqrt{I_jI_{j+1}}\cos(\phi_j-\phi_{j+1})-\frac{\beta U}{2}\sum_j I_j^2+\beta \mu \sum_j I_j\bigg),
\end{equation}
where $d\vec{I}d\vec{\phi}$ indicates integration over all variables and $\beta$ is the inverse temperature. We first perform the integration over the phases:
\begin{equation}
   \label{int} Z_a=(2\pi)^{L}\int d\vec{I}\exp\bigg(\beta\mu\sum_jI_j-\frac{\beta U}{2}\sum_jI_j^2\bigg)\prod_{j=1}^{L-1}\mathcal{I}_0\left(2\beta J\sqrt{I_jI_{j+1}}\right),
\end{equation}
where $\mathcal{I}_0$ is the modified Bessel function of zeroth order. This result follows from the known integral:
\begin{equation}
    \int_0^{2\pi}d\phi_1\exp\left(\cos\left(\phi_1-\phi_2\right)\right)=2\pi\mathcal{I}_0(1).
\end{equation}
Exact integration over the numbers $I_j$ in (\ref{int}) can only be done in the $U\to \infty$ limit, representing the Bessel functions as a power series. The result is given by:
\begin{equation}
    Z_a=(2\pi)^L \exp\bigg(\frac{\beta\mu^2L}{2U}\bigg)\Big(\frac{1}{2\beta U}\Big)^{\frac{L}{2}}\sum_{k_1,k_2,\ldots, k_{L-1}=0}^{\infty}\frac{(2\beta J^2U^{-1})^K}{\prod_{j=1}^{L-1}(k_j!)^2}\prod_{i=0}^{L-1}\Gamma\Big(\frac{k_i+k_{i+1}+1}{2}\Big).~~~~~~~~~\label{zfull}
\end{equation}
In the above we have denoted $K\equiv\sum_{i=1}^{L-1} k_i$, $k_0 \equiv 0, k_L \equiv 0$. This is essentially an expansion in $\beta J^2/U$, therefore although the annealed approximation would be expected to work only in the superfluid regime where the numbers evolve faster than in the Mott regime (though still slower than the phases), in fact we can also write a controlled expansion which remains valid into the Mott regime too. In this regime it makes sense to only keep the zeroth and first term in the above expansion:
\begin{equation}
Z_a=(2\pi)^L\exp\bigg(\frac{\beta\mu^2L}{2U}\bigg)\bigg(\frac{\pi}{2\beta U}\bigg)^{\frac{L}{2}-1}\bigg(\frac{\pi}{2\beta U}+ (L-1)\frac{J^2}{U^2}\bigg).\label{partfun1}
\end{equation}
From the above relation we can directly obtain the thermodynamic energy and the heat capacity:
\begin{equation}
E_a=\frac{L}{2}\left(T-\frac{\mu^2}{U}\right),~~C_a=\frac{L}{2}.\label{annealedec}
\end{equation}
Therefore, in the annealed regime the system behaves thermodynamically as an ideal gas -- the expressions (\ref{annealedec}) are just the consequence of the equipartition theorem, with the twist that the degrees of freedom are not enumerated by particles but by sites (each site contributes exactly $\frac{1}{2}$ to the internal energy). In this case the normal diffusion regime (which we observe at very late times) is naturally expected. For later comparison to the quenched case, we also give the free energy in the annealed regime (also at the first order in the expansion (\ref{zfull})):
\begin{equation}
   \label{annealed} F_a=-\frac{1}{\beta}\bigg[L \log2\pi+\frac{\beta\mu^2L}{U}+\frac{L}{2}\log\bigg(\frac{\pi}{2\beta U}\bigg)+\frac{2(L-1)J^2\beta}{\pi U}\bigg].
\end{equation}

\subsection{Quenched free energy}\label{subsubsec2}

For the quenched calculation we only integrate over the phases, calculate the logarithm of the outcome (i.e., the free energy) and then average it over the number variables, arriving at the following integral:
\begin{equation}
F_q=-\frac{1}{\beta}\int d\vec{I}\log\left[\left(2\pi\right)^L\prod_{j=1}^{L}\sum_{k=0}^\infty\frac{(\beta^2J^2)^k(I_jI_{j+1})^k}{(k!)^2}\exp\left(\beta\mu I_j-\frac{\beta U I_j^2}{2}\right)\right].\label{partfunquenched}
\end{equation}
This integral can be computed exactly, even when $N$ does not go to infinity. This is done by introducing the hyperspherical coordinates. We present only the final result:
\begin{eqnarray}
    F_q&=&-\frac{1}{\beta L!}\bigg[L\log 2\pi+\frac{\beta\mu L}{L+1}-\frac{\beta U L}{(L+1)(L+2)} + \nonumber\\ &+&L!(L-1)\sum_{n=1}^{\infty}\frac{(-1)^{n+1}}{n}\sum_{k_1,\ldots k_{n}=1}^{\infty}\frac{(\beta^2J^2)^{K_n}}{\prod_{j=1}^{n}(k_j!)^2}\frac{\Gamma^2(K_n+1)}{\Gamma(2K_n+L+1)}\bigg].\label{fquenchfull}
\end{eqnarray}
Above we define $K_n\equiv \sum_{i=1}^n k_i$ (for $K$ as previously defined in (\ref{zfull}) we thus have $K=K_L$). For some intuitive insight we only take the first term in this expansion, i.e. the term with $n=1, k_1=1$:\footnote{The radius of convergence of equation (\ref{fquenchfull}) is quite difficult to analyze, as the last term in the sum behaves as a product of Bessel $\mathcal{I}_0$ functions suppressed by a polynomial with a leading term being proportional to $K_n^{L+1/2}$. We expect it to diverge for a wide range of the $\beta J$ parameter. When $\beta J \ll 1$, we have a controlled approximation leading to results in equations (\ref{partfunquenched1}) and (\ref{quenchedec}). }
\begin{equation}
    F_q=-\frac{1}{\beta L!}\bigg[L\log 2\pi+\frac{\beta^2J^2(L-1)}{(L+1)(L+2)}+\frac{\beta\mu L}{L+1}-\frac{\beta U L}{(L+1)(L+2)}\bigg].\label{partfunquenched1}
\end{equation}
From here we derive the internal energy and heat capacity:
\begin{equation}
E_q=\frac{U-\mu L-2\beta J^2}{L!L},~~C_q\sim\frac{\beta^2J^2}{L!L}.\label{quenchedec}
\end{equation}
We see a few surprising things in this analysis. First, there is the factor of $L!$ in the quenched quantities which appears from the integration over the $L$-dimensional sphere (which does not happen in the annealed case). Because of this the strict thermodynamic limit $L\to\infty$ predicts zero free energy, i.e. the breakdown of thermodynamics. This basically means that in the quenched regime there is no thermal equilibrium.

Assuming that the above results make some sense for finite $L$, we note that the heat capacity now decreases with temperature as $1/T^2$, which is unexpectedly consistent with the results of \cite{rizzatti2020double}, obtained without assuming the semiclassical approximation. Therefore, the quenching of the slow variables essentially mimics the influence of quantum corrections. This can be understood as a manifestation of the truncated Wigner approximation (TWA) \cite{Polkovnik2003,Polkovnik2003Main,PolkovnikSciPost,PolkovnikAnnals} where classical equations of motion, averaged over the initial conditions, reproduce the leading quantum (Bogolyubov) behavior.

To conclude, the thermodynamic analysis teach us the following:
\begin{enumerate}
    \item The annealed (number and phase) partition function behaves as an ideal gas, suggesting that individual orbits undergo random walk and hence normal diffusion. This holds in any dimension and for any parameter regime.
    \item The quenched (number only) partition function is ill-defined in the thermodynamic limit, suggesting a lack of thermodynamic equilibrium on timescales short compared to the number (action) evolution. The result at finite particle number reproduces the leading quantum contribution to heat capacity, in analogy with the fact that averaging classical orbits over some initial distribution reproduces the leading quantum behavior as in TWA.
\end{enumerate}

\section{Path integral approach}

Since both annealed and quenched free energies capture only special limits, it is useful to try interpolating between them. In principle this requires the full path integral formalism, but of course this is impossible to do in practice (except as a costly numerical computation). We will just include the leading (quadratic) correction to the quenched regime, reducing the dynamics to linearized oscillations around the equilibrium points.

It will turn out that taking into account even small linearized oscillations of the number variables already bridges the gap between the annealed and the quenched regime. The Lagrangian of the system is given by Legendre transformation of the Hamiltonian:
\begin{eqnarray}
    L&=&-H+\sum_i \dot{\phi_i}I_i=\sum_i\left[2J\sqrt{I_iI_{i+1}}\cos(\phi_i-\phi_{i+1})-\frac{U}{2}I_i^2+\mu I_i\right]+\nonumber\\
    &+&\sum_i\left[-\mu I_i +UI_i^2-J\sqrt{I_iI_{i+1}}\cos(\phi_i-\phi_{i+1})-J\sqrt{I_iI_{i-1}}\cos(\phi_i-\phi_{i-1})\right]=\nonumber\\
    &=&\frac{U}{2}\sum_i I_i^2.
\end{eqnarray}
The last equality follows from relabeling of the summation index $i\mapsto i+1$ and the fact that cosine is an even function. The resulting Lagrangian is formally given by the sum of "kinetic energies" of non-interacting point particles of mass $1/U$. One might be baffled that this Lagrangian does not contain any angles except for those from the beginning and end of the chain. But this is a consequence of the constrained nature of the system: the constraint of number conservation $\sum_i I_i=1$ upon resolving (i.e. writing for example $I_L=1-\sum_{i=1}^{L-1}I_i$) mixes all numbers with the phase $\phi_L$, which in turn mixes with $\phi_{L-1}$ and so on. Moreover, the Lagrangian only depends on the Coulomb repulsion $U$, and not on $\mu$, so we have a non-interacting Lagrangian with a single scale.\footnote{Of course, the canonical transformations from the original variables to $(I_j,\phi_j)$ depend on both $U$ and $\mu$ hence the system in fact depends on all the parameters as it has to be.} This also holds for the rectangular and cubic lattice. Now we find the Euclidean action and the partition function (where we do not explicitly write out the normalization $\mathcal{N}$):
\begin{equation}
    \mathcal{S}_0=\frac{U}{2}\int_{t_1}^{t_2}dt\sum_j I_j(t)^2,~~\mathcal{Z}(\beta)=\mathcal{N}\int \mathcal{D}[\vec{q}]\exp(-\mathcal{S}_0),
\end{equation}
where $q$ are the variables describing the dynamics of the system around some equilibrium point (below we define $\vec{q}$ in detail). The first approximation for the number $I_i$ is $I_i=\mathrm{const.}$ Therefore we can define the positions $x(I_i)$ so that $\ddot{x}(I_i)=0$. The path integral is now calculated by perturbing the classical trajectory by some $\Vec{\eta}$. It can easily be shown that
\begin{equation}
    \mathcal{S}_0[x_\mathrm{cl}(I_i)+\eta_i]=\mathcal{S}_0[x_\mathrm{cl}(I_i)]+\mathcal{S}_0[\eta_i].
\end{equation}
%Knowing the partition function for a noninteracting particle
%\begin{equation}
%     \mathcal{Z}_1=V_d \bigg(\frac{m}{2\pi\beta}\bigg)^{d/2},
%\end{equation}
%\begin{equation}
%    \mathcal{Z}_i=\frac{V_L}{\sqrt{2\pi\beta U}}
%\end{equation}
Substituting $q_i(t)=x_\mathrm{cl}\left(I_i\left(t\right)\right)+\eta_i(t)$ the integral becomes:
\begin{equation}
    \mathcal{Z}(\beta)=\mathcal{N}\int \mathcal{D}[\vec{\eta}] \exp\left(-\sum_i \mathcal{S}_{0i}\right)=\mathcal{N}\prod_i\int d\eta_i(t)\exp(-\mathcal{S}_{0i})=\mathcal{N}\prod_i \mathcal{Z}_i.
\end{equation}
In the first approximation the result is:
\begin{equation}
    \mathcal{Z}(\beta)=\mathcal{N}\bigg(\frac{V_L^2}{2\pi\beta U}\bigg)^{L/2}=\mathcal{N}\bigg(\frac{2\pi}{\beta U}\bigg)^{L/2}=Z_a^{(0)}.
\end{equation}
The outcome (and the thermodynamics) is the same as for the annealed partition function! However, the starting assumptions and thus the interpretation are not the same:
\begin{enumerate}
\item{The annealed partition function is obtained by integrating the statistical weights with the full nonlinear Hamiltonian over the whole phase space, essentially assuming that both the numbers and the phases are ergodic and explore the whole phase space during their evolution, for any $U$.}
\item{On the other hand, in the path integral calculation we introduce just the leading, linear correction to the opposite, quenched regime as we model the evolution of the numbers by linear oscillations around fixed positions -- and we have a formally noninteracting Lagrangian. Of course, the interactions are hidden in the constraints between the variables.}
\end{enumerate}
Therefore, there is a tradeoff: ergodic/annealed dynamics for the full interacting system captures the same thermodynamic regime as the near-quenched dynamics but in the linearized approximation.

We would not expect that the annealed approximation for the thermodynamics coincides already with the leading correction to the quenched linear approximation. But this result encapsulates the puzzling behavior of transport from section \ref{sec3}: even in the strongly chaotic regime the early-time transport behaves strongly anomalously, i.e. shows strong correlations and is strongly non-Markovian. And likewise, even in the weakly chaotic regime we eventually reach the normal diffusive regime at very late times for many initial conditions. Here we have expressed the question sharply in thermodynamic terms and hope to answer it in the future.

%The annealed partition function tells us that each site contributes $\frac{1}{2}$ to the heat capacity, however the sites do not have mass, whilst the path integral tells us that in the first approximation, we also have $L$ quasiparticles but now with mass $1/U$ and whose momenta are given by their action values. The path integral formalism for the trajectories of constant $I$ sees the ideal gas setup. We know that actions are not constant but the higher terms are much more complicated for calculation. 

\subsection{Mean kinetic energy}

In order to gain some more intuition for the thermodynamic behavior we derive the dispersion relation for the mean kinetic energy $\bar{\epsilon}$. In other words we calculate the second moment of the angular velocity $\dot{\phi}_l$, which dominates the kinetic energy over $\dot{I}_l$ which tends to be smaller. The calculation is straightforward:
\begin{equation}
    \langle\dot{\phi}^2_l\rangle=\frac{1}{Z_a}\int d\vec{I} d\vec{\phi}\dot{\phi_l}^2 \exp\bigg(2\beta J\sum_{j=1}^{L-1}\sqrt{I_jI_{j+1}}\cos(\phi_j-\phi_{j+1})-\frac{\beta U}{2}\sum_j I_j^2+\beta \mu \sum_j I_j\bigg).
\end{equation}
When the number of sites is large we arrive at:
\begin{equation}
\bar{\epsilon}=\frac{1}{2U}\langle\dot{\phi}_l^2\rangle=\frac{\mu^2}{2U}-\frac{\mu}{2U}\Bar{v}+\frac{T}{2},
\end{equation}
where $\Bar{v}$ is the average velocity of ideal gas at temperature $T$. Although the partition function formally corresponds to the ideal gas, the dispersion relation is modified, due to the implicit interactions/constraints involved in the definition of phase-number (action-angle) variables.

%Indeed when $U \to\infty$ the equipartition theorem result holds and the particles, whatever they are, each hold $1/2$ of the heat capacity.
%What is also of interest the dispersion relation also depends linearly on the mean momentum, as $\Bar{p}=\Bar{v}/U$.

\section{Discussion and conclusions}\label{sec13}

The key result of our work on chaos in the Bose-Hubbard model are the integer superdiffusion coefficients, which are present at least until some late time. This regime is strongly non-ergodic and is expected to be closer to the quenched regime for the thermodynamics. The fact that in this regime the free energy is not even defined (i.e., finite) except for very high temperatures is not surprising: it merely indicates that the system is very far from equilibrium and does not have a meaningful thermodynamics description, except when thermal averaging becomes strong enough to overcome the non-ergodicity from regular islands. The possibility to reach the annealed regime from the minimally perturbed quenched regime suggests that one should be able to find that a unified approach describing both the anomalous and normal regime.

The normal diffusion regime is \emph{a priori} easier to understand. The system equilibrates and, according to our annealed free energy, it is effectively described as a diffusing ideal gas, with some minimal modification.

One question for further work is if we always reach the normal diffusion regime, or there are quasi-invariant structures which always preclude normal diffusion in some cases? Another task is to include quantum corrections, which will likely reveal new physics. 

Finally, it is crucial to know to what extent our specific conclusions depend on the details of the Bose-Hubbard Hamiltonian. The same general dynamical picture -- that of mixed phase space with both regular and chaotic components, when the number of degrees of freedom is not too large -- was found for example in the Dicke model \cite{Villasenor:2022spy,Meier:PhysRevE.107.054202}. Even deep in the quantum regime and in the presence of dissipation, a similar spectrum of behaviors from regular to chaotic, with strong dependence on initial conditions, was found \cite{Kolovsky:PhysRevA.100.013610,Muraev:2023,Ferrari:2023zfe}. What has not been done is to explore what happens when the number of degrees of freedom is of order a few dozen or even a few hundred, which we have checked explicitly in this work and \cite{markovic2023chaos}. Therefore, we can wonder if the same kind of anomalous transport exists in Dicke and other models. At least in the semiclassical limit this is not difficult to check and we plan to address this issue.

\backmatter

\bmhead{Acknowledgments}

We are grateful to Marco Schiro, Jak\v{s}a Vu\v{c}i\v{c}evi\'c, Filippo Ferrari, Fabrizio Minganti, Zlatko Papi\'c and Andrea Richaud for stimulating discussions. Work at the Institute of Physics is funded by the Ministry of Education, Science and Technological Development and by the Science Fund of the Republic of Serbia. M.~\v{C}. would like to acknowledge the Mainz Institute for Theoretical Physics (MITP) of the Cluster of Excellence PRISMA+ (Project ID 39083149) for hospitality and partial support during the completion of this work.

\bibliography{sn-bibliography}
\end{document}